\documentclass[twocolumn,floatfix,aps,prl,showpacs]{revtex4}
\usepackage{graphicx}
\newcommand{\beq}{\begin{equation}}
\newcommand{\eeq}{\end{equation}}
\newcommand{\beqa}{\begin{eqnarray}}
\newcommand{\eeqa}{\end{eqnarray}}
\newcommand{\vep}{\varepsilon}

\begin{document}

\title{Spin selective Aharonov-Bohm oscillations in a lateral triple quantum dot}

\author{F. Delgado$^{1,2}$, Y.-P. Shim$^1$, M. Korkusinski$^1$, 
L. Gaudreau$^{1,3}$, S. A. Studenikin$^1$, A. S. Sachrajda$^1$, and P. Hawrylak$^{1,2}$} 

\affiliation{$^1$ Institute for Microstructural Sciences,  
       National Research Council, Ottawa, Ontario, Canada K1A 0R6} 
\affiliation{$^2$ Department of Physics, University of Ottawa, 
Ottawa, Ontario, Canada K1N 6N5}   

\affiliation{$^3$ R\'egroupement Qu\'eb\'ecois sur les Mat\'eriaux de
  Pointe, Universit\'e de Sherbrooke, Qu\'ebec, Canada J1K 2R1}

\begin{abstract}  
We present a theory for spin selective Aharonov-Bohm oscillations in a lateral triple quantum dot. We show that to understand the Aharonov-Bohm (AB) effect in an interacting electron system
within a triple quantum dot molecule (TQD) where the dots lie in a ring 
configuration requires one to not only consider electron charge but also
spin. Using a Hubbard model supported by microscopic calculations we show that,
by localizing a single electron spin in one of the dots, the current through the TQD molecule
 depends not only on the flux but also on the relative orientation of the spin of the
  incoming and localized electrons.
AB oscillations are predicted only for the spin singlet electron complex resulting in a magnetic field tunable ``spin valve''.
\end{abstract}  
  

\maketitle

%
%
The Aharonov-Bohm\cite{aharonov_bohm_pr1959} (AB) effect 
results from the accumulation of phase by a charged particle moving in a ring threaded by a magnetic flux~\cite{buttiker_imry_pra1984,webb_washburn_prl1985}.
AB oscillations are detected e.g. in the magnetization of 
a macroscopic number of electrons in metallic rings\cite{levy_dolan_prl1990}
as well as in the optical emission from a charged exciton in a nanosize
semiconductor quantum ring~\cite{bayer_korkusinski_prl2003}.
The preparation, manipulation and detection of individual spins of localized electrons
in nanoscale semiconductor systems are important elements of nano-spintronic applications~\cite{awschalom_loss_snt2002,zutic_fabian_rmp2004,
sachrajda_hawrylak_book2003}, with  
efficient generation and detection of spin polarized carriers playing a crucial role.
The electron spins can be localized in single and coupled semiconductor quantum dots (QDs)
defined and controlled electrostatically\cite{ciorga_sachrajda_prb2000,koppens_buizert_nature2006,
petta_johnson_science2005,vidan_westervelt_jsupercond2005,gaudreau_studenikin_prl2006,ihn_sigrist_njphys2007}
with potential applications as elements of electron-spin based
  circuits~\cite{brum_hawrylak_sm1997,loss_divincenzo_pra1998}, coded
   qubits~\cite{hawrylak_korkusinski_ssc2005},
entanglers~\cite{saraga_loss_prl2003},
rectifiers  and ratchets~\cite{stopa_prl2002,vidan_westervelt_apl2004}.
The Spin blockade technique in a double dot system used for the conversion of spin to charge information has played an important 
role in the development of such applications~\cite{ono_austing_science2002}.
%

\begin{figure}
\begin{center}
\includegraphics[angle=0,height=0.95\linewidth,width=0.95\linewidth]{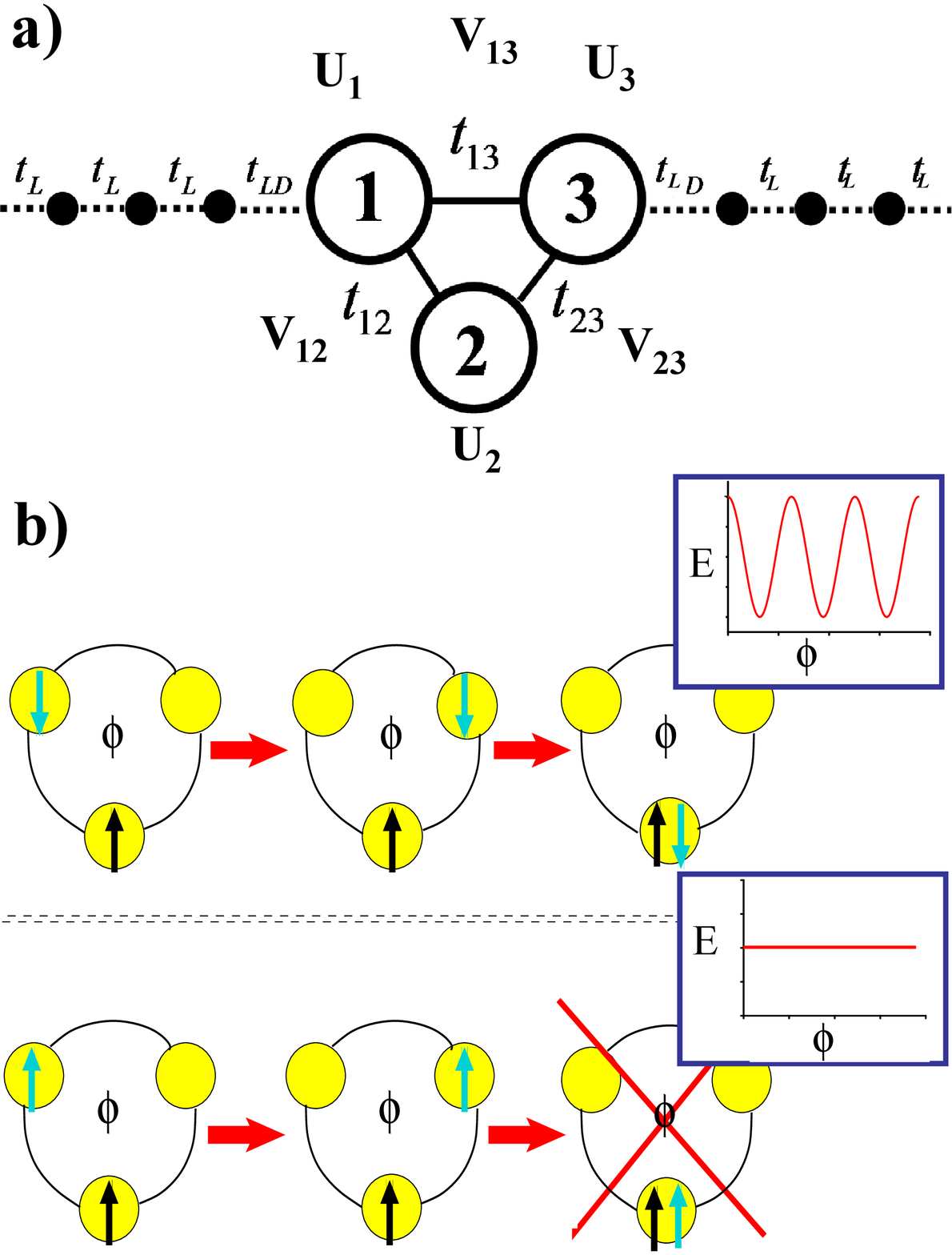}
\end{center}
\caption{{\bf (a)} Schematic diagram of the TQD
close to the considered QP.
{\bf (b)}
Electrons with antiparallel spins can form a loop an the corresponding energy levels
$E$ experiment AB like oscillations with the magnetic flux $\phi$ while
electrons with parallel spin are spin-blockaded.}
\label{fig1}
\end{figure}

The possibility of the co-existence of spin blockade
with AB oscillations in a lateral TQD in a ring 
 geometry  \cite{gaudreau_studenikin_prl2006,vidan_westervelt_jsupercond2005,ihn_sigrist_njphys2007}
was discussed in Ref. \onlinecite{delgado_shim_prb2007}. In this paper we describe a TQD,
shown schematically in Fig.~\ref{fig1}(a), where two dots, 1 and 3,
 are connected to the leads and in addition to dot number 2.
A single electron spin is localized in dot 2 by lowering the confining potential. 
The transport of an additional electron through the TQD will now depend on
 the relative orientation of the spin of the incoming and localized electrons.
  If the two spins are anti-parallel, as shown in Fig.~\ref{fig1}(b), 
  the additional electron can tunnel from the left lead to dot 1, and proceed
 either directly to dot 3 or through dot 2 to dot 3 and thus to the right lead.
  In the presence of the magnetic
  field the two paths acquire a different phase and can interfere, resulting in the AB oscillations 
  of the current amplitude (upper inset).
    When the spin of the incoming electron is parallel to the spin
   of the electron in dot 2, the Pauli exclusion principle prevents tunneling through dot 2, resulting in 
   a single tunneling path and the absence of AB oscillations.   (lower inset).
We present here the theory of 
these spin selective AB oscillations in transport through a TQD
in a perpendicular magnetic field
 with a controlled number of electrons.  
The electronic properties of a TQD are treated  by a fully microscopic LCHO-CI
 approach\cite{gimenez_korkusinski_prb2007} and by Hubbard and t-J models with  exact
many-electron eigenstates obtained using
configuration-interaction (CI) method~\cite{korkusinski_gimenez_prb2007}.
The Fermi Golden Rule and the sequential tunneling
approach\cite{muralidharan_datta_prb2007} are used
to calculate the current
through the TQD weakly connected to two non-interacting leads. The current flows when the chemical potential of the TQD  is equal to the chemical potential of the leads. This can also be understood in terms of degeneracies of many electron charge configurations $(N_1,N_2,N_3)$, with $N_i$ the number of electrons in dot $i$. 
The degeneracy point described here, referred to as the quadrupole point (QP),
involves the one electron 
configuration $(0,1,0)$ and two-electron configurations $(1,1,0)$, $(0,2,0)$ and
$(0,1,1)$, with one electron always confined in dot $2$, as shown in Fig. \ref{fig1}(b).

%
%
%
For clarity we only present results of the Hubbard model with one orbital per dot~\cite{korkusinski_gimenez_prb2007,delgado_shim_prb2007}.
The Hamiltonian of the TQD subject to a uniform perpendicular magnetic field,
${\bf B}=B {\bf \hat{z}}$, is given by
\beqa
  H&=&
   \sum\limits_{i,\sigma}  E_{i\sigma} d_{i\sigma}^{\dag} d_{i\sigma}
+ \sum\limits_{\sigma,i,j;\;i\neq j}
             \tilde{t}_{ij} d_{i\sigma}^{\dag} d_{j\sigma}\cr
&&+ \sum\limits_{i} U_{i}n_{i\downarrow} n_{i\uparrow}
+ {1\over{2}}\sum_{i,j;\; i\neq j}V_{ij}
          \varrho_{i} \varrho_{j},
\label{htqd}
\eeqa
where the operators $d_{i\sigma}$ ($d_{i\sigma}^\dag$) annihilate (create)
an electron  with spin $\sigma=\pm 1/2$ on orbital $i$ ($i=1,2,3$).
$n_{i\sigma} = d_{i\sigma}^\dag d_{i\sigma}$ and $\varrho_{i} = n_{i\downarrow} + n_{i\uparrow}$
are the spin and charge density on orbital level $i$.
Each dot is represented by a single orbital with energy 
$E_{i\sigma}= E_i+ g^* \mu_B B \sigma+E_0$, where
$g^*$ is the effective Land\'e $g$-factor, $\mu_B$ is the Bohr magneton and
$E_0$ is the common energy shift of the three dots
 measured from the Fermi level of the leads which is
tunable by external gates. 
The dots are connected by magnetic field dependent hopping matrix elements
$\tilde{t}_{ij}=t_{ij}e^{2\pi i \phi_{ij}}\;$~\cite{peierls_zphys1933}.
For the three dots in an equilateral configuration
$\phi_{12}=\phi_{23}=\phi_{31}=-\phi/3$ and  $\phi_{ji}=-\phi_{ij}$,
where $\phi=BA/\phi_0$ is the number of magnetic
flux quanta threading the area $A$ of the triangle,
$\phi_0=hc/e$ is the magnetic flux quantum, $e$ is the electron charge,
$c$ is the speed of light and $\hbar$ is the Planck's constant.
The interacting part of the Hamiltonian is parametrized by the on-site Coulomb
repulsion, $U_{i}$, and the interdot direct repulsion term $V_{ij}$.

%
%
In order to describe transport through the TQD we first 
determine the QP of the isolated TQD.
We start by determining the ``classical QP''
where we neglect the inter-dot tunneling and require the four configurations
 $A\equiv (1,1,0)$, $B\equiv(0,2,0)$, $C\equiv(0,1,1)$ and
 $D\equiv(0,1,0)$ to have equal energy. Their energies are
 $\epsilon_A=E_1+E_2+V_{12}+2E_0$, $\epsilon_B=2E_2+U_{2}+2E_0$,
 $\epsilon_C=E_2+E_3+V_{23}+2E_0$, and $\epsilon_D=E_2+E_0$. 
The QP condition without tunneling requires
$\epsilon_A=\epsilon_B=\epsilon_C=\epsilon_D+\mu_L$, where $\mu_L$ is the chemical
potential of the leads.
This implies that at the QP 
$E_1^Q=\mu_L-E_0-V_{12}$, $E_2^Q=\mu_L-E_0-U_{2}$ and $E_3^Q=\mu_L-E_0-V_{23}$.

\begin{figure}
\begin{center}
\includegraphics[angle=0,height=0.85\linewidth,width=0.9\linewidth]{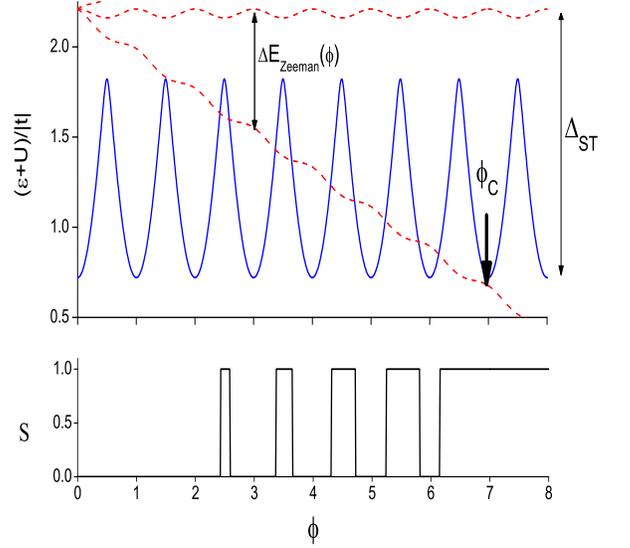}
\end{center}
\caption{Lowest energy spectrum  of the two electron TQD 
at the QP (upper panel) and total spin of the ground state (lower panel) versus the magnetic
flux. The QP condition was found numerically for $\delta_1=\delta_3=2.44|t|$ and $\delta_2=2.77|t|$.
}
\label{fig3}
\end{figure}

Let us consider now the case of finite tunneling matrix elements.
The $(0,1,0)$ charge configuration describes the two spin states
of an electron localized in dot $2$, 
$|2\sigma\rangle \equiv d_{2\sigma}^\dag|0\rangle$ with energy
$E_{2}$  ($|0\rangle$ is the vacuum state).
%
%
The two electron classical charge configurations $(1,1,0)$, $(0,2,0)$ and
$(0,1,1)$ correspond to the following quantum spin singlet configurations:
$|S_1\rangle =  \frac{1}{\sqrt{2}}( d_{1\uparrow}^{\dag}d_{2\downarrow}^{\dag}
+ d_{2\uparrow}^{\dag}d_{1\downarrow}^{\dag})|0\rangle$,
$|S_2\rangle =  d_{2\uparrow}^{\dag}d_{2\downarrow}^{\dag}|0\rangle$ and
$|S_3\rangle =  \frac{1}{\sqrt{2}}( d_{2\uparrow}^{\dag}d_{3\downarrow}^{\dag}
+ d_{3\uparrow}^{\dag}d_{2\downarrow}^{\dag})|0\rangle$.
The Hamiltonian describing the motion of the 
spin singlet pair takes the form
\beqa
\hat{H}_{S}=\left(\begin{array}{ccc}
\epsilon_A & \sqrt{2}t_{12}e^{-2 \pi i\phi/3}  & t_{13}e^{2\pi i\phi/3} \\
\sqrt{2}t_{12}^*e^{2 \pi i\phi/3} &\epsilon_{B} &  \sqrt{2}t_{23}e^{-2 \pi i\phi/3} \\
 t_{13}^*e^{-2\pi i\phi/3} &  \sqrt{2}t_{23}^*e^{2 \pi i\phi/3} & \epsilon_{C} \\
\end{array}
\right).
\label{hsin}
\nonumber
\eeqa
At the classical QP, we have $\epsilon_A=\epsilon_B=\epsilon_C$.
If $t_{23}=t_{12}=\sqrt{2}t_{13}$, we can diagonalize the Hamiltonian exactly by 
Fourier transforming into a new basis:  
$|{K_1}\rangle =1/\sqrt{3}\left(|1\rangle+|2\rangle  
+|3\rangle\right)$,   
$|{K_2 }\rangle =1/\sqrt{3}\left(|1\rangle  
+e^{i 2\pi/3}|2\rangle+e^{i 4\pi/3}|3\rangle\right)$ and
$|{ K_3}\rangle =1/\sqrt{3}\left(|1\rangle  
+e^{-i2\pi/3}|2\rangle+e^{-i4\pi/3}|3\rangle\right)$    
with eigenvalues $\vep_1=E-2|t|\cos\left(2\pi\phi/3\right)$,  
$\vep_2=E-2|t|\cos\left[2\pi(\phi+1)/3\right]$ and  
$\vep_3=E-2|t|\cos\left[2\pi(\phi-1)/3\right]$, respectively. 
Since one of the electrons is kept in dot $2$, the energy spectrum of a pair of singlet electrons
is essentially the same as that of a  single electron added to a resonant TQD,
with the energy levels oscillating with a period of one flux quantum~\cite{delgado_shim_prb2007}.
Away from the resonance the level crossing is replaced by anti-crossing.


A pair of spin triplet electrons describes only $(1,1,0)$ and
$(0,1,1)$ charge configurations. The corresponding two spin triplet configurations 
for $S_Z=1$ are 
$|T_1\rangle=d_{1\uparrow}^{\dag}d_{2\uparrow}^{\dag}|0\rangle$ with energy
$\epsilon_A(B)$
and $|T_2\rangle=d_{2\uparrow}^{\dag}d_{3\uparrow}^{\dag}|0\rangle$ with energy
$\epsilon_C(B)$, with $\epsilon_A(B)=\epsilon_A+g^*\mu_B B S_z$.
The eigenenergies of the $2\times 2$ triplet Hamiltonian are 
$\vep_{T}^{\pm}=1/2\big\{\epsilon_A(B)+\epsilon_C(B)
\pm\left[\left(\epsilon_A(B)-\epsilon_C(B)\right)^2
+4|t_{13}|^2\right]^{1/2}\big\}$.
By comparing the singlet and triplet eigenvalues we see that singlet is the ground state at $B=0$ 
and the eigenvalues of the triplet do not oscillate as a function of the magnetic field.
Even at this qualitative level, we obtain a remarkable result
that triplet states do not oscillate with the magnetic field while singlets do.


In the case of finite tunneling each classical configuration is no longer an eigenstate of the system.
Therefore, we will define QP as
the point in the parameter space where the
ground state energies of two and one electrons differ by $\mu_L$ and the three
degenerate two-electron configurations are found with the same probabilities.
Then, at the QP $E_i=E_i^Q+\delta_i$,
where the energies $\delta_i$ are quantum corrections that are obtained numerically,
with $\delta_1=\delta_3$ for the symmetric case described here.

\begin{figure}
\begin{center}
\includegraphics[angle=0,height=1.\linewidth,width=1.\linewidth]{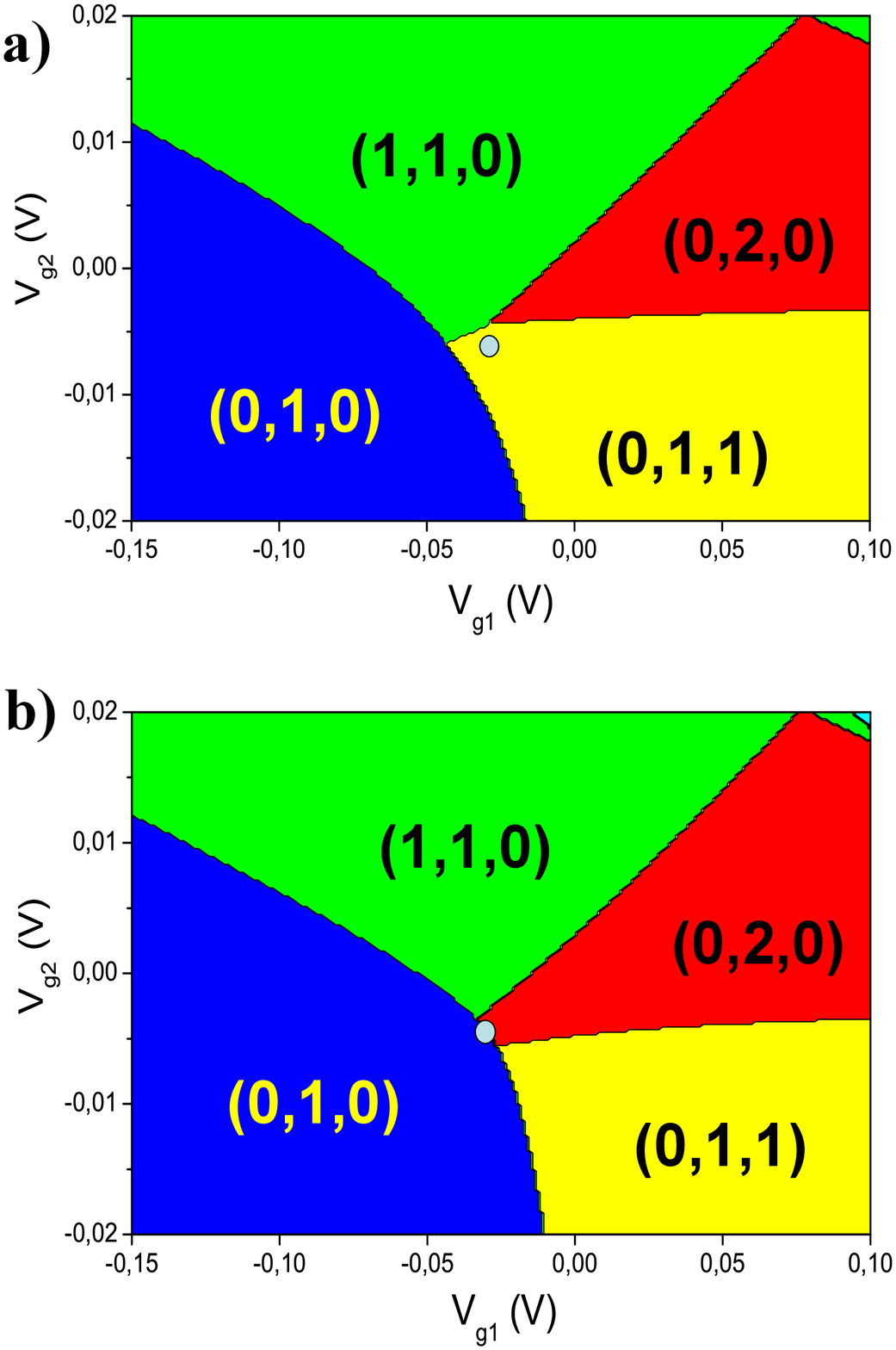}
\end{center}
\caption{Stability diagram of the TQD close to the QP
with charge configurations ${\bf (1,1,0),\;(0,2,0)}$, ${\bf (0,1,1)}$ and ${\bf (0,1,0)}$
at
{\bf (a)} $\phi=0$ and
{\bf (b)}, $\phi=0.44$.
 The classical QP ($t=0$) is found at $V_{g1}=V_{g2}=0$ while
the quantum one at $\phi=0.44$ is indicated by the white circle. 
}
\label{figsta}
\end{figure}

%
%
We shall analyze now the magnetic field dependence of the two electron energy spectrum 
close to the QP.
The numerical CI calculations include the full Hilbert space generated from
the three orbital levels. 
Hubbard parameters were obtained from the LCHO calculation 
with an interdot distance of $61.2$ nm: $t=-0.23$ meV,
$U_i=50|t|$ and $V_{ij}=10|t|$. $g^*=-0.44$ corresponding to GaAs will be assumed.
The upper panel of Fig.~\ref{fig3} shows the lower part of the energy spectrum for
$E_0=-|t|$, while the lowest panel indicates the total spin of the ground state.
%
%
%
As was described above, the singlet (solid line) is the ground state at $B=0$. 
The lowest energy of a singlet  oscillates with period of one flux quantum
while the energy of a triplet decreases monotonically with increasing magnetic field due to Zeeman energy. 
Notice that triplets show a small oscillation due to a coupling with higher energy configurations.
The oscillating singlet energy and monotonically decreasing triplet energy 
leads to a number of transitions between singlet and triplet
with increasing magnetic field. These transitions interrupt the AB oscillations of the singlet, 
and lead to their end at a critical value of the magnetic field,
 $B_C=-\Delta_{ST}/g^*\mu_B$ with $\phi_C=AB_C/\phi_0$,
 indicated in Fig.~\ref{fig3}.
 Above $\phi_C$ the triplet is the ground state.
Hence the presence of a trapped electron should lead to AB oscillations of the tunneling electron,
interrupted and eventually terminated by the singlet-triplet transitions.

Figure \ref{figsta} shows the dominant charge ground state configurations of the TQD
at two different values of the magnetic flux quantum, $\phi=0$ (upper panel)
and $\phi=0.44$ (lower panel), versus the voltages $V_{g1}$ and $V_{g2}$ for $E_0=-|t|$.
Here it has been
assumed that the on-site energies $E_i$'s change 
 linearly with the voltages $V_{g1}$ and
$V_{g2}$, $E_i=\alpha_i V_{g1}+\beta_i V_{g2}+\gamma_i$,
with $\alpha_i,\;\;\beta_i$ extracted from 
experiment in Ref.~\onlinecite{gaudreau_studenikin_prl2006}. 
For the chosen value of $E_0$, the
zero magnetic field
stability diagram shows only a triple point, while at $\phi$ the QP is clearly visible.


%
We now turn to the illustration how these spin selective AB oscillations can be 
observed in transport experiment. 
Following Ref.~\onlinecite{delgado_hawrylak_jphy2008},
the Hamiltonian of the TQD connected to two leads is given by $H=H_{L}+H_{TQD}+H_{LD}$,
where $H_{L}$ is the Hamiltonian describing the two non-interacting leads,
$H_{TQD}$ corresponds to the isolated triple dot where we 
assume that the on-site energies change with the applied bias $\Delta V$ as
$E_{i\sigma}\to E_{i\sigma}-\Delta V/2$ and
$H_{LD}$ is the tunneling Hamiltonian between the leads and the TQD.
The leads are described with a one-dimensional tight-binding model
with nearest neighbor hopping $t_L$,
on-site energies $\epsilon_{L}$ and $\epsilon_{R}$ for the left (right) leads
and coupling strength between dots and leads $t_{LD}$\cite{delgado_hawrylak_jphy2008}.
The current through the system is evaluated using a set of master equations for the
occupation probabilities within the sequential tunneling
approximation~\cite{muralidharan_datta_prb2007}.
In this approach we neglect higher order processes such as cotunneling
which is important for high tunnel-coupling strengths and for temperatures below the
Kondo temperature~\cite{ingersent_ludwig_prl2005,kuzmenko_kikoin_prl2006}.
The occupation probabilities are then calculated using a detailed balance condition
imposed by the conservation of charge.
The spin components of the current in the linear regime
at the lowest order in the coupling $t_{LD}$ and at zero temperature
are then given by
$I^\sigma=
e\pi/(2\hbar)|t_{LD}|^2\rho(\vep_F)\Delta V C^\sigma \delta(\vep_F-(\vep_{2,G}-\vep_{1,G}))
$
where $\vep_{2,G}\; (\vep_{1,G})$ is the ground state energy of the two (one) electrons and
 $\rho(\vep_F)$ is the density of states in the leads 
at the Fermi level. Here we make the assumption 
 $\rho_L(\vep_{F,L})\approx \rho_R(\vep_{F,R})$.
$C^\sigma=1/3$ for $\sigma=\downarrow$ (singlet ground state) and
 $C^\sigma=1$ for $\sigma=\uparrow$ (triplet ground state).

\begin{figure}
\begin{center}
\includegraphics[angle=0,width=1.1\linewidth]{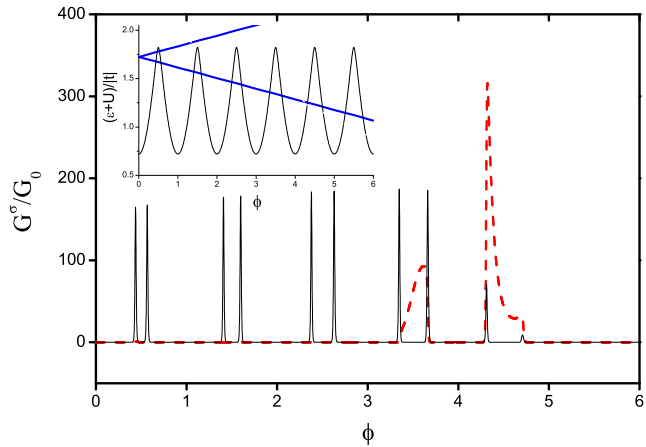}
\end{center}
\caption{Spin down (solid line) and spin up (dashed line) components of the conductance 
in units of $G_0'=e^2|t_{LD}|^2/\hbar|t_L|^2$
versus the number of flux quantum $\phi$ for the same parameters
as in Fig. \ref{fig3}.
The inset shows
the energy spectrum of the two electron complex (as in Fig. \ref{fig3}), 
together with the one electron lowest levels (thick blue line).}
\label{fig4}
\end{figure}

Next we present the results for the 
linear conductance $G=I/\Delta V$.
The calculations were done at $50$ mK ($k_B T=0.0145|t|$),
$\Delta V= 2\times10^{-3}|t|$ and 
$\mu_L=0$.
In addition, $|t_L|=23.72$ meV $\gg |t|,\; E_0,\;\Delta V$.
Since transport through the TQD is allowed whenever the single-particle ground state
and the two-particle ground state are on resonance,
the AB oscillations of the energy spectra lead to repeated peaks in current.
The spin components of the conductance $G^\sigma=I^\sigma/\Delta V$
are shown in Fig.~\ref{fig4}.
At low magnetic fields, the spin down current is dominant
and transport is mainly through the lowest oscillating singlet state.
When the ground state of two particles becomes triplet, spin
up current is dominant until the current is totally suppressed.



In summary, the presence of an extra electron localized in one dot 
of a ring-like TQD 
leads to spin selective AB oscillations as a function of magnetic field.
The energy of the singlet
 ground state oscillates as a result of the interference between the two possible paths
while the triplet state does not oscillate since one of the paths is spin blockaded
by the presence of a localized particle.
The magnetic field orients the spin of the localized particle leading to
the transport of electrons
with a specific spin polarization.
The AB oscillation of the singlet electron pair is reflected as peaks in 
the spin-down polarized current.
At higher magnetic field, the Zeeman energy causes a singlet-triplet transition,
which results in a change of the dominant spin component of the current.

%
The Authors acknowledge support by the QuantumWorks Network 
and the Canadian Institute for Advanced Research.


\end{document}